\documentclass[amsmath,amssymb,aps,pra,twocolumn,showpacs,nofootinbib]{revtex4-1}

\usepackage{graphicx}
\usepackage[colorlinks=true,linkcolor=blue,citecolor=blue,urlcolor=blue]{hyperref}
\usepackage{color}

\newcommand{\om}{\omega}
\newcommand{\Om}{\Omega}
\newcommand{\ii}{i}
\newcommand{\ee}{e}
\newcommand{\dd}{d}
\newcommand{\dx}{\dd x}

\newcommand{\ommin}{\om_{\rm min}}
\newcommand{\ommax}{\om_{\rm max}}

\begin{document}

\title{Spontaneous quantum emission from analog white holes in a nonlinear optical medium}

\author{Stefano Finazzi}
\email{stefano.finazzi@univ-paris-diderot.fr}
\altaffiliation[Currently at ]{Laboratoire Mat\'eriaux et Ph\'enom\`enes Quantiques, Universit\'e Paris Diderot-Paris 7 and CNRS, B\^atiment Condorcet, 10 rue Alice Domon et L\'eonie Duquet, 75205 Paris Cedex 13, France.}
\affiliation{INO-CNR BEC Center and Dipartimento di Fisica, Universit\`a di Trento, via Sommarive 14, 38123 Povo--Trento, Italy}
\author{Iacopo Carusotto}
\email{carusott@science.unitn.it}
\affiliation{INO-CNR BEC Center and Dipartimento di Fisica, Universit\`a di Trento, via Sommarive 14, 38123 Povo--Trento, Italy}

\begin{abstract}
We use a microscopic quantum optical model to compute the spectrum of quantum vacuum emission from strong laser pulses propagating in nonlinear optical media. Similarities and differences with respect to the emission of analog white holes as predicted by quantum field theory in curved spacetime are highlighted. Conceptual issues related to the role played by the material dispersion and to the presence or absence of the horizon are clarified. Critical comparison with available experimental data is made.
\end{abstract}

\pacs{%
      42.65.-k,
      04.62.+v,
      42.50.Lc 
}
\date{\today}
\maketitle

\section{Introduction}

The most celebrated example of spontaneous particle creation from vacuum fluctuations was predicted by Hawking~\cite{hawkingnat,hawking75} in the context of quantum field theory in curved spacetime and consists of the emission of a thermal radiation from the horizon of black holes. In the last decades, the extreme difficulty of detecting this emission from astrophysical black holes has stimulated the investigation of analogous phenomena in condensed-matter or optical systems~\cite{lr}. 

The first claim of observation of spontaneous analog Hawking radiation in a laboratory was indeed made in~\cite{faccioexp,faccionjp}: following the proposal in~\cite{ulf_science}, a strong infrared pump pulse was sent through a nonlinear medium (fused silica in the quoted experiment) and created a moving modulation of the refractive index. As a result, the speed of optical photons inside (outside) the pulse is smaller (larger) than the pulse velocity: Seen from the reference frame comoving with the pulse, the leading (trailing) edge of the pulse appears then as the analog of a black (white) hole horizon. Even if the interpretation of the experimental results as Hawking radiation is still considered as controversial by some authors~\cite{comment,reply}, alternative explanations are based on simplified models~\cite{angus,unruhuniverse}.

A first step toward the construction of a complete theory of these phenomena was made in~\cite{kinetics,gradino}, where a microscopic one-dimensional model of this system was developed and used to predict the emission spectrum for black-hole configurations. 
The purpose of this article is to apply the model of~\cite{gradino} to compute the spectrum of the spontaneously emitted radiation by the analog white horizon at the trailing edge of the pulse, from which most of the quantum vacuum radiation is expected to be emitted in actual experiments: As a consequence of nonlinear effects during propagation, this edge quickly becomes substantially steeper than the leading one. As a first step, we consider the idealized case of a large refractive index modulation and we highlight the consequences of the multibranched Sellmeier dispersion of fused silica as compared to the simpler dispersion of nonpolar dielectrics such as diamond, for which the analogy with quantum field theory on curved spacetime is accurate. We then move to more realistic cases inspired by the experimental parameters~\cite{faccioexp,faccionjp}. Our results in this respect will hopefully not only contribute to the on-going debate on the interpretation of the experiment~\cite{comment,reply,angus,unruhuniverse}, but also shine light on the conceptual issues related to the role of the material dispersion and of the horizon on the quantum vacuum radiation process.

\section{Theoretical model}
\label{sec:theory}

Following~\cite{gradino}, we describe matter-light interaction in terms of a generalized Hopfield model~\cite{hopfield} where the electromagnetic field is coupled to several matter polarization fields. These are modeled as uniformly distributed charged harmonic oscillators with elastic constants $\beta_i$ and frequencies $\Om_i$ (see Appendix~\ref{app:equations}). 
With suitable choices of the parameters, this model is able to reproduce the correct Sellmeier dispersion~\cite{sellmeier}
\begin{equation}\label{eq:sellmeier}
 c^2 K^2 = \Om^2 +\sum_{i=1}^n\frac{4\pi\beta_i\, \Om^2}{1-\Om^2/\Om_i^2}
\end{equation}
of transparent dielectrics. For fused silica, $n=3$ resonances are needed with $\beta_{1,2,3}=0.07142,\,0.03246$, and $0.05540$ and $\hbar\Om_{1,2,3}=0.1253,\,10.67$, and $18.13\,\mbox{eV}$~\cite{refractiveindex}.

\begin{figure*}
 \includegraphics[width=\textwidth]{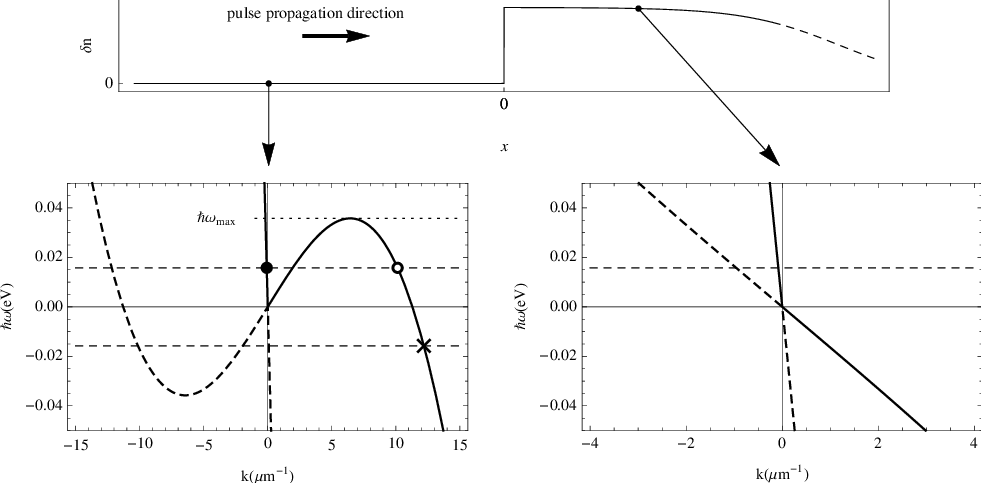}
  \caption{\label{fig:simpledispersion}Top panel: Sketch of the refractive index profile for a right-going pulse generating an analog white-hole horizon at $x=0$.
Bottom row: single-resonance dispersion relation seen from the comoving reference frame in the external (left) and internal (right) regions. Solid (dashed) curves represent positive (negative) norm branches. Parameters of the single resonance: $\Om_0=\Om_2$, $\beta_0=\beta_2$, $\epsilon=0.3$, and $v=0.83c$.}
\end{figure*}

As it is sketched in the top panel of Fig.~\ref{fig:simpledispersion}, the increase of the refractive index caused by the strong pump pulse is described as a sudden change of the local values of $\beta_i$ and $\Om_i$. 
For the sake of simplicity, we restrict ourselves to a simplified one-dimensional geometry which already contains all the basic features of the physics under investigation. Extension to the full three-dimensional case introduces severe technical difficulties which go far beyond the scope of the present work.\footnote{For a pump pulse with plane-parallel edges, the transverse wavevector $\mathbf{k}_\parallel$ along the plane is conserved: for each $\mathbf{k}_\parallel$, the scattering matrix remains finite dimensional, with just a very complex geometrical structure of modes~\cite{kinetics}. The situation is far more complicated if one wishes to describe the scattering and amplification of electromagnetic zero-point fluctuations on a bullet-shaped pulse with a nontrivial profile along the transverse directions. Finally, for the model to be complete one should also include the temporal evolution of the pulse under the combined effects of nonlinearity and dispersion while it propagates through the medium. A first study of the new features that may arise in this case is given in~\cite{angus}.}
Inspired by the experiments, we also focus our attention on the steep trailing edge (located at $X=vT$) of a pulse propagating rightward at speed $v>0$: This is expected to be an accurate approximation since light scattering and quantum vacuum emission from the much smoother leading edge are generally negligible.
Outside the pulse (i.e., for $X<vT$), $\beta_i$ and $\Om_i$ coincide with the tabulated values of the material, while inside the pulse (i.e., for $X>vt$) we take
\begin{equation}\label{eq:betaomleft}
 \beta'_{i}=(1+\epsilon)\beta_{i},\qquad \Om'_{i}=(1+\epsilon)^{-1/2}\Om_{i}.
\end{equation}
In the actual calculations, the value of $\epsilon$ is chosen in a way to give the desired value of the refractive index change $\delta n$ for photons in the optical window. 
In experiments, $\delta n$ is controlled through the intensity of the pump pulse via $\delta n=n_2\,I$, $n_2$ being the optical Kerr nonlinearity of the material. In the case of fused silica, $n_2\approx3\times10^{-16}\,\mbox{cm}^2/\mbox{W}$, which means that an intensity $I\approx3\times10^{12}\,\mbox{W}/\mbox{cm}^2$ is required to get $\delta n\approx 0.001$~\cite{faccioexp}. Larger $\delta n$ are limited by the damage threshold of the material.

On the other hand, the velocity of the pulse $v$ can be tuned either by using pump lasers of different wavelengths and therefore different group velocities $v_0$ or by focusing the pump beam with a conic lens. While a standard Gaussian pulse propagates in vacuum at velocity $c$, the highest intensity region of a Bessel pulse~\cite{durninprl,gori,mcdonald} obtained by focusing such a pulse through a conic lens travels at an effective superluminal velocity $v=c_{\delta}=c/\cos\theta>c$. The so-called Bessel angle $\theta$ is controlled through the Snell law $\sin(\gamma + \theta) = n_{\rm lens} \sin\gamma$ by the aperture $\pi-2\gamma$ of the conic lens and by its refractive index $n_{\rm lens}$~\cite{hermanwiggins}. Analogously, the effective velocity $v=v_{\delta}=v_0/\cos\delta>v_0$ of a Bessel pulse propagating in a medium is higher than the group velocity $v_{0}$ and depends on the Bessel angle $\delta$ measured in the medium~\cite{faccionjp}.

\begin{figure*}
 \includegraphics[width=\textwidth]{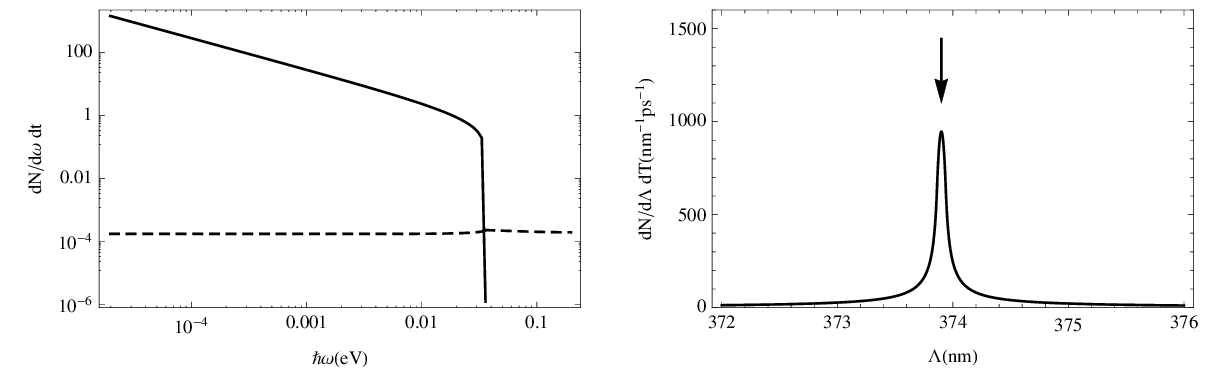}
  \caption{\label{fig:simpleflux}
Emission spectrum from an analog white hole in a nonpolar dielectric such as diamond. Left panel: comoving frame emission spectrum for the two positive-norm outgoing modes: The solid (dashed) line corresponds to the mode indicated by an open (filled) dot on the dispersion shown in Fig.~\ref{fig:simpledispersion}. Right panel: total emission spectrum in the laboratory frame. The arrow indicates the wavelength at which the phase velocity equals the pulse velocity, i.e., $\omega=0$. Same parameters as in Fig.~\ref{fig:simpledispersion}.}
\end{figure*}

Assuming that the pulse shape does not appreciably change during propagation, the system can be described in the frame comoving with the pulse at velocity $v$ by a time independent Hamiltonian (throughout this article, quantities measured in the laboratory vs. comoving frame are denoted by upper vs. lower case symbols). The field equations can then be solved in Fourier space with respect to time: within each homogeneous region on either side of the pulse edge, the dispersion is given by the Lorentz transform of the Sellmeier dispersion Eq.~\eqref{eq:sellmeier}. Matching conditions have then to be imposed at the pulse edge. This procedure (more details can be found in the Appendix and in Ref.~\cite{gradino}) provides the scattering matrix relating incoming (i.e., directed toward the pulse edge, as seen from the comoving frame) and outgoing modes. As this matrix mixes modes with positive and negative norms at a given comoving frequency $\omega$, destruction and creation operators are correspondingly mixed. As usual, this leads to a finite intensity of spontaneous quantum vacuum emission even for an initial vacuum state with no incoming particles. Once the emission spectrum in the comoving frame is known for all modes, the one in the laboratory frame is obtained by a Lorentz transformation of the emission rate for each outgoing mode contributing to a given laboratory frame frequency $\Omega$ and then by summing over all the modes.\footnote{Most remarkably, whereas fluxes and emission rates coincide in the comoving frame where the pulse is stationary, a difference between them arises in the laboratory frame where the source---the pulse---is moving at velocity $v$. This simple fact may have some importance when comparing different calculations and/or experimental data. In this article, inspired from the detection scheme used in the experiments~\cite{faccioexp,faccionjp}, we have chosen to describe the spectral properties of the laboratory frame emission in terms of the emission rates, that is, the number of photons emitted per unit time. The total number of emitted photons per pulse is then obtained by multiplying the emission rate by the time it takes the pulse to propagate along the nonlinear crystal.}

\section{Results}

\subsection{Simplified dispersion relation}
\label{subsec:simpledisp}

As a first application, we study the case of a nonpolar dielectric such as diamond, where the only resonances in the Sellmeier dispersion Eq.~\eqref{eq:sellmeier} are in the UV and the dispersion is regular from $\Om=0$, where it is almost linear with a low-frequency refractive index $n_0$, all the way through the IR and the visible range. Without loss of generality, the calculations have been performed with a single UV resonance at $\Omega=\Omega_2$ and $\beta=\beta_2$.

In particular, the pulse speed $v$ is chosen $c/(n_0+\delta n_0)<v<c/n_0$ in order to have an analog white-hole horizon. The dispersion relation of the lower polariton branch seen in the comoving frame is plotted in the two panels on the bottom row of Fig.~\ref{fig:simpledispersion} for, respectively, the external (left) and internal (right) region; the upper polariton branch lies in the UV far outside the field of view and plays no role in the physics under examination. Solid (dashed) lines indicate positive (negative) norm modes. The white-hole nature of this configuration is visible in that no light can propagate from the horizon into the internal region. The fact that in-going modes with both positive and negative norms exist at all frequencies is responsible for the emission of quantum vacuum radiation at all frequencies.

The emission spectra on positive norm modes with positive frequency $\omega>0$ in the comoving frame are shown in the left panel of Fig.~\ref{fig:simpleflux}. The solid line corresponds to the Hawking-like emission in the mode indicated as an open dot on the dispersion curve shown in the left panel of Fig.~\ref{fig:simpledispersion}: as expected on the basis of the analogy with quantum field theory on curved spacetimes, its spectrum displays the usual thermal-like $1/\om$ divergence at low frequencies~\cite{lr,carlos,scottreview}.
As the dispersion of the outgoing Hawking mode (open circle in the upper left panel) does not extend beyond its maximum value $\ommax$, the Hawking emission then disappears at high frequencies $\om>\ommax$. This effect was originally discussed for black-hole configurations in generic dispersive systems in~\cite{macherRP1} and then illustrated in the nonlinear optical context in~\cite{gradino}.

\begin{figure*}
 \includegraphics[width=\textwidth]{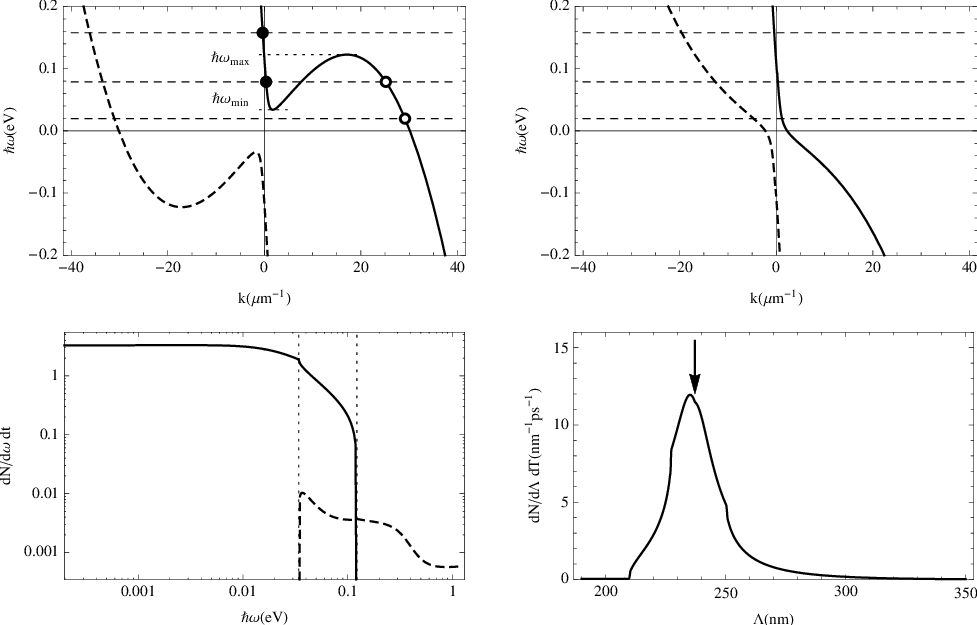}
 \caption{\label{fig:optical}Upper panels: Optical branch ($\Om_1<\Om<\Om_2$) of the dispersion relation of fused silica [Eq.~\eqref{eq:sellmeier}] as seen from the comoving frame. The left (right) panels refer to the external (internal) region of an analog white hole.  Solid (dashed) curves represent positive (negative) norm branches. Parameters: $v=0.66 c$ and $\epsilon=0.3$, giving a strong refractive index modulation $\delta n\approx 0.12$ in the visible range.
  Lower left panel: Emission spectrum into the optical branch as seen from the comoving frame. The solid (dashed) line refers to the emission into the positive norm modes indicated by open (filled) dots in the dispersion shown in the upper left panel. The emission on the other upper and lower branch is much weaker.
  Bottom right panel: Total emission spectrum in the laboratory frame. The arrow indicates the wavelength $\Lambda_0$ at which the phase velocity equals the pulse velocity, i.e., $\omega=0$.
  }
\end{figure*}

On the other hand, the intensity of the emission on the second outgoing mode (filled dot on the dispersion curve in the left panel of Fig.~\ref{fig:simpledispersion}), which propagates backward with respect to the pulse in the laboratory frame, is nonzero for all values of $\om$ but much weaker, showing that the spontaneous radiation occurs almost only on forward-propagating modes.
By energy conservation in the comoving frame, the emission on the outgoing positive-norm negative-frequency mode indicated by a cross on the dispersion curve in the left panel of Fig.~\ref{fig:simpledispersion} is given by the sum of the two emissions on the positive norm modes. As a consequence, the spectrum measured in the comoving frame is symmetric for $\omega\to-\omega$.

The right panel of Fig.~\ref{fig:simpleflux} shows the total emission spectrum as observed in the laboratory frame. The thermal peak of the Hawking radiation around $\om\simeq0$ in the comoving frame is Doppler shifted in the laboratory frame to a finite $\Om=|\gamma v k_0|$ with $k_0$ defined by the intersection of the dispersion relation with the $\om=0$ line. In physical terms, $k_0$ corresponds to the wavelength $\Lambda_0$ at which the phase velocity of light equals the pulse velocity and is indicated in the figure by the vertical arrow. Note that wavelengths $\Lambda>\Lambda_{0}$ ($\Lambda<\Lambda_{0}$) correspond to positive (negative) comoving frequencies $\omega$. We have checked that this result is general and holds irrespective of the relative magnitude $\epsilon$ of the refractive index change across the analog white-hole horizon. A similar result was obtained in~\cite{faccionuovo}.

\subsection{Full Sellmeier dispersion}

We now proceed to study the complete Sellmeier dispersion Eq.~\eqref{eq:sellmeier} of fused silica where the features of quantum vacuum emission are expected to be dramatically modified by the presence of a low-frequency pole in the infrared region of the material dispersion, corresponding to an optical phonon transition. We start by investigating the case of a large refractive index jump $\delta n=0.12$. Even though reaching this value in fused silica would perhaps require a very high intensity well beyond the damage threshold, we consider this possibly unrealistic case for a conceptual reason: This allows us to disentangle the two different effects due, respectively, to the presence of the extra pole in the dispersion relation and to the tiny value of the refractive index jump $\delta n$. Indeed, as we shall see in Sec.~\ref{subsec:weak}, in realistic experimental situations they both contribute to distort the thermal spectrum found in the previous section.

The three resonances at $\Om_{i=1,2,3}$ split the dispersion relation into four polariton branches. The highest one at $\Om>\Om_3$ is never involved in the physics under investigation. In what follows, we shall refer to the other three as lower ($\Om<\Om_1$), optical ($\Om_1<\Om<\Om_2$), and upper ($\Om_2<\Om<\Om_3$) branches.

While our calculations fully take into account all branches, in Fig.~\ref{fig:optical} we focus our attention on the optical branch only. For a pump pulse on the optical branch~\cite{kinetics,gradino}, the main difference as compared to Fig.~\ref{fig:simpledispersion} is the presence of a forbidden gap in the dispersion shown in the upper left panel for the region outside the pulse. The facts that the analog white-hole horizon is active only in a finite frequency window $(\ommin,\ommax)$ and there are no low-momentum modes below a certain frequency $\ommin$ may naively appear as minor differences, but they have a dramatic effect on the quantum vacuum emission as it removes two of the four solutions which are involved in the Hawking emission process at low $\om$. 

In particular, the emission spectra plotted as a solid line in the lower left panel of Fig.~\ref{fig:optical} no longer show the typical $1/\om$ thermal-like divergence of Hawking emission, but the total emission goes to a constant value as $\omega$ tends to zero, in agreement with the results of~\cite{Scott_thesis,ulfscott}, where this behavior was firstly described.
As the analog white-hole horizon is present only in the frequency range $(\ommin,\ommax)$, the only remaining signature of the Hawking effect is the rapid growth of the emission for decreasing $\om$ within this window. 
On the other hand, the emission on the backward-propagating mode (filled dots in the upper left panel) and on the lower and upper branches is much smaller at all frequencies; see the dashed line spectrum in the lower-left panel.

Another remarkable consequence of the infrared pole is a slight spectral shift of the emission in the laboratory frame: in contrast to the standard Hawking case of Fig.~\ref{fig:simpledispersion}, the peak of the emission spectrum in the laboratory frame (lower right panel) is in fact slightly shifted from the wavelength $\Lambda_0$ indicated by an arrow where the pulse velocity equals the phase velocity.

\subsection{Weak refractive index modulations}
\label{subsec:weak}

As a last example, we now investigate the experimentally most relevant case of a weak refractive index modulation.
In this case, the presence of an analog horizon requires a very fine tuning of the pulse velocity. However, 
as it was noticed in~\cite{gradino} for the analog black-hole case, negative norm modes with positive comoving frequency are present at all speeds and quantum vacuum radiation is emitted irrespective of the presence or absence of an analog horizon.

\begin{figure}
\includegraphics[width=\columnwidth]{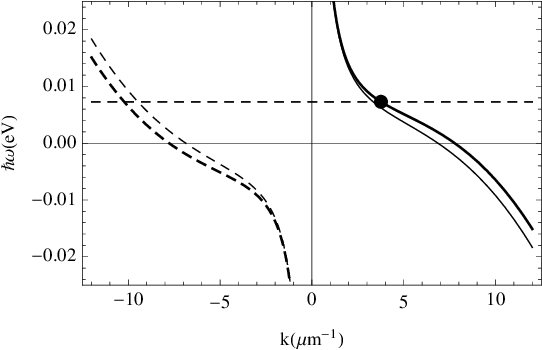}
 \caption{\label{fig:dispersion_smalln} Optical branch ($\Om_1<\Om<\Om_2$) of the dispersion relation of fused silica [Eq.~\eqref{eq:sellmeier}] as seen from the comoving frame for a Bessel pulse with $v=0.6885c=v_{\delta=6.7^\circ}$ and small refractive index modulation of $\delta n\approx 0.0016$. The thick (thin) lines refer to the external (internal) region of an analog white hole.  Solid (dashed) curves represent positive (negative) norm branches.}
\end{figure}

This important fact is illustrated in Fig.~\ref{fig:dispersion_smalln}, where the dispersion relation in the comoving frame is plotted for a typical pulse velocity of $v=0.6885c$ which can be experimentally achieved by a Bessel pulse with angle $\delta=6.7^{\circ}$.\footnote{\label{foot:daniele}This is the measured experimental value of the parameter $\delta$ used in the experiment of~\cite{faccioexp,faccionjp}. Note that this value differs slightly but significantly from the nominal value of $7^\circ$ quoted in the experimental papers~\cite{faccioexp,faccionjp} [D. Faccio, (private communication)].} For this value of the pulse velocity, no horizon is present. Furthermore, the weakness of the refractive index jump ($\delta n\approx 0.0016$ in the figure) makes the dispersion relations in the internal (thick lines) and external (thin lines) regions very similar.

\begin{figure}
\includegraphics[width=\columnwidth]{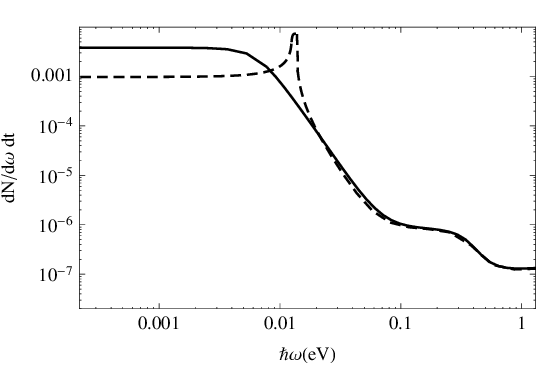}
\caption{\label{fig:fluxcom}Solid line: Emission spectrum in the comoving frame into the positive norm mode indicated by the dot in the dispersion relation of Fig.~\ref{fig:dispersion_smalln}, for a horizonless geometry formed by a Bessel pulse with $v=0.6885c=v_{\delta=6.7^\circ}$. Dashed line: Sum of the emission spectra into the positive norm mode indicated by the closed and open dots in the dispersion relation of the upper left panel of Fig.~\ref{fig:optical} for a geometry with white-hole horizon formed by a Gaussian pulse with $v=v_{0}\approx0.684c$. In both configurations $\delta n\approx 0.0016$, corresponding to an input energy of $E=1280\,\mu\mbox{J}$ in the experiment of Ref.~\cite{faccioexp}.}
\end{figure}

This yields an extremely faint spontaneous emission on the unique outgoing positive norm mode (closed dot in the dispersion shown in Fig.~\ref{fig:dispersion_smalln}), whose flux is represented in the comoving frame by the solid line in Fig.~\ref{fig:fluxcom}: the emission is quickly suppressed at large $\om$ and tends to a constant in the $\om\to0$ limit. In the absence of a horizon determining $\ommin$ and $\ommax$, the extension of the low-frequency plateau is related to the position of the inflection point of the dispersion. As compared to Fig.~\ref{fig:optical}, the much weaker emission intensity is a consequence of the small value of $\delta n$.

\begin{figure*}[htbp]
 \includegraphics[width=\textwidth]{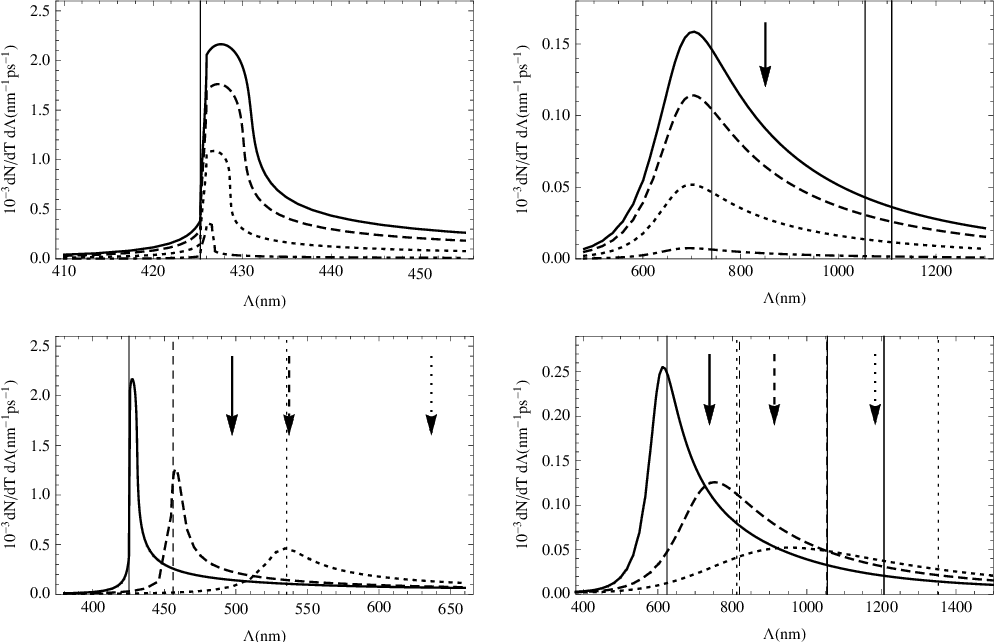}
 \caption{\label{fig:spectra} Laboratory frame emission spectra in the case of a weak refractive index modulation obtained with a 1055~nm-laser pulse. 
 Upper panels: Gaussian pulse propagating at $v=v_0\approx0.684c$ (left) and Bessel pulse propagating at $v=0.6885c= v_{\delta=6.7^\circ}$ (right). The different curves refer to different values of the refractive index modulation $\delta n\approx1.6,1.3,0.88,0.33\,\times 10^{-3}$ (solid, dashed, dotted, dot-dashed line), corresponding to input energies of $E=1280,\,1080,\,720,\,270\,\mu\mbox{J}$ in the experiment of Ref.~\cite{faccioexp}.
 Lower panels: fixed value of $\delta n=1.6\times10^{-3}$ and different pulse velocities. Left: $v=0.684c, 0.685c, 0.686c$ (solid, dashed, dotted line, corresponding to $\delta=0$ Gaussian and $\delta=3,5$ Bessel pulses). Right: $v=0.688c, 0.689c, 0.690c$ (solid, dashed, dotted line, corresponding to $\delta=6,7,8$ Bessel pulses).
 In each panel, the arrows indicate the wavelength $\Lambda_0$ at which the phase velocity equals the pulse velocity. 
 The vertical lines indicate the wavelengths $\Lambda_l^{(i)}$ at which resonant radiation can occur~\cite{faccioneg}. The line style coding follows the one of the spectra. In the upper panels, $\Lambda_0$ and $\Lambda_l^{(i)}$ are the same for all curves. In the left upper panel, $\Lambda_0=497$~nm and the corresponding arrow falls outside the field of view.}
\end{figure*}

The dashed curve in Fig.~\ref{fig:fluxcom} represents the sum of the fluxes of the emission on the two outgoing positive norm modes indicated by open and closed dots, respectively, on the dispersion relation in the upper left panel of Fig.~\ref{fig:optical}, for a horizon configuration with the same values of $\delta n\approx 0.0016$.
Comparison of the solid and dashed curves confirms the results of~\cite{gradino} for the black-hole case, namely, that for such weak values of $\delta n$ the comoving frame emission spectrum is not very much affected by the presence or absence of the horizon. The only (quantitatively minor) difference is the peak in the tiny frequency window where the Hawking-like emission from the horizon is present.

The emission spectra seen from the laboratory frame are studied in Fig.~\ref{fig:spectra}. Inspired by the experiments in~\cite{faccioexp,faccionjp}, we consider a pump pulse of wavelength $\lambda=1055$~nm, corresponding to a group velocity of $v_0=0.684\,c$.
The upper left panel illustrates the case of a Gaussian pulse for which $v=v_0$, while the upper right panel illustrates the case of a Bessel pulse propagating at the faster speed $v=0.6885\,c$ corresponding to a $\delta=6.7^\circ$ Bessel angle (see footnote~\ref{foot:daniele}). In each panel, four different values of the refractive index modulation are considered.
For the slower velocity $v=v_0$ (upper left panel), an analog white-hole horizon is present for all considered values of~$\delta n$ and the dispersion has the same shape as shown in the upper panels of Fig.~\ref{fig:optical}. On the other hand, for the faster velocity $v=v_{\delta=6.7^\circ}$~(upper right panel), the dispersion has the qualitative shape shown in Fig.~\ref{fig:dispersion_smalln} with no horizon.

The most noteworthy feature of these spectra consists in a significant spectral shift of the emission peak from the wavelength $\Lambda_0$ at which the pulse velocity equals the phase velocity. In contrast to the nonpolar dielectric case where the emission peak is exactly located at $\Lambda_0$ (Fig.~\ref{fig:simpleflux}), the presence of the IR pole is responsible for a shift of the emission peak. While in the large modulation case of Fig.~\ref{fig:optical} this shift was a minor correction, it can be quite large for a weak modulation: In Fig.~\ref{fig:spectra}, this fact is visible by comparing the position of the emission peaks with the arrows indicating $\Lambda_0$.

Two effects can be invoked to justify this shift. On the one hand, the presence of the IR pole eliminates the $1/\om$ divergence of the comoving frame emission, which is replaced by a finite constant value in the $\om\to 0$ limit (see Figs.~\ref{fig:optical} and~\ref{fig:fluxcom}). On the other hand, even if the spectrum measured in the comoving frame is symmetric for $\omega\to-\omega$ (see Sec.~\ref{subsec:simpledisp}), its shape is strongly distorted by the Doppler transformation from the comoving to the laboratory frame: This transformation is a combination~\cite{gradino} of the Lorentz transformation of the frequency differential as $d\omega = \gamma(1-v/V_{g})d\Omega$, where $V_{g}$ is the photon group velocity measured in the laboratory frame, and a relativistic time dilation $\Delta T =\gamma \Delta t$,
\begin{equation}
 \frac{dN}{d\Omega\,dT} = \left|1-\frac{v}{V_{g}}\right| \frac{dN}{d\omega \,dt}.
\label{eq:Doppler}
\end{equation}
According to this expression, the laboratory frame emission is strongly suppressed at those frequencies for which the group velocity $V_{g}$ is close to the pulse speed $v$, that is, when the group velocity $v_{g}$ measured in the comoving frame is close to zero. 
The shift of the emission peak observed in Fig.~\ref{fig:spectra} is thus explained by the fact that the minimum of $v_{g}$ indicated by the inflection point in the dispersion relation of Fig.~\ref{fig:dispersion_smalln} is typically located at low positive values of $\omega$. This breaks the symmetry $\omega\to-\omega$ and, as a result, the spectrum measured in the laboratory frame is peaked at $\Lambda<\Lambda_{0}$.

\section{Comparison with experimental data}

To complete our discussion, it is interesting to critically compare our theoretical prediction with available experimental data~\cite{faccioexp,faccionjp}.
The first puzzling feature of the experiment is the observation of emission at $90^\circ$ with respect to the propagation direction of the pulse.
Based on purely kinematic arguments~\cite{kinetics}, the spontaneous Hawking emission cannot exit a perfectly homogeneous dielectric in this direction. To justify the observation, some scattering process has to be invoked, caused for instance by impurities of the glassy fused silica medium~\footnote{This possibility is not excluded by the authors of~\cite{faccioexp,faccionjp} (private communication).}.
Under the reasonable assumption that the scattering is wavelength independent, one can attempt to compare our predictions with the results of Figs.~3 and 4 of~\cite{faccioexp}.

For Gaussian pulses, the agreement of the predicted position of the peak of the emission with the experimental data in Fig.~4 of~\cite{faccioexp} is quite good: This fact is even more remarkable if one notices that, in this case, the peak emission wavelength around $\lambda=400$~nm is significantly different from the wavelength $\Lambda_0=497$~nm at which the phase velocity equals the pulse velocity. Moreover, the broader experimental spectrum can be explained by the variation of the pulse velocity during propagation.

For Bessel pulses, the agreement is only at the qualitative level and severe quantitative differences are visible. First, from our theoretical calculations, one would expect a much wider emission spectrum than the experimentally observed one: So far, no explanation for this discrepancy is available. Second, the peak emission is theoretically expected at a wavelength around $\lambda=700$~nm, significantly shorter with respect to the experimental one of $\lambda=850\div 875$~nm reported in Fig.~3 of~\cite{faccioexp}. 

While one must not forget that important deviations might occur from the one-dimensional approximation of our theory, another explanation for this shift may come from the strong sensitivity of the spectral features on the exact value of the pulse velocity $v$, experimentally controlled by the Bessel angle $\delta$. This fact is illustrated in the lower panels of Fig.~\ref{fig:spectra}, showing the emission spectrum for various velocities ranging from $v=0.684c$ to $0.690c$ at a fixed value of $\delta n=1.6\times10^{-3}$: For instance, a small change of $\delta$ from $6$ to $7^\circ$ already leads to a 100~nm-shift of the emission peak.

Furthermore, when analyzing the emission spectra, one must keep in mind a crucial result of this work, namely, that in polar dielectrics such as fused silica the emission peak for small modulations $\delta n$ is significantly shifted from the wavelength $\Lambda_{0}$ corresponding to the comoving frame zero frequency. This is due to the presence of a pole in the infrared domain, which restricts Hawking processes to a finite frequency window and makes the comoving frame emission spectrum tend to a constant for $\om\to 0$ (instead of diverging as $1/\om$ as would be naively expected from the gravitational analogy). As a consequence of the Jacobian in the Doppler transform Eq.~\eqref{eq:Doppler}, the emission peak in the laboratory frame is then found at a shorter wavelength than $\Lambda_0$. In the figure, this feature is clearly visible by comparing the position of the peaks with the vertical arrows indicating the position of $\Lambda_0$ for the parameters of each spectrum. It is worth noting that slightly different results were obtained in Ref.~\cite{faccionuovo} using an alternative numerical technique based on the numerical solution of Maxwell equations. 

Finally, it must not be forgotten that other effects may produce similar spectra possibly with stronger intensity, e.g., the resonant radiation (RR) process and its ``negative'' counterpart (NRR) recently observed in~\cite{faccioneg}. By ``resonant radiation''\footnote{Note that the same expression is sometimes used in a broader sense to indicate various optical processes involving the negative norm branch of the dispersion~\cite{NRR_sc_rep}.} we indicate here a scattering process where two pump photons interact via the nonlinearity. One of them falls back in a stimulated way into the pump wave packet, while the other one acquires a different frequency. Energy conservation in the comoving frame~\cite{faccioneg} determines the wavelengths allowed by this scattering process, which have to share the same comoving frame frequency of the laser pump, irrespective of the norm of the mode. 

In Fig.~\ref{fig:spectra}, the laboratory wavelengths at which resonant radiation processes (both positive and negative ones) can occur are indicated by straight vertical lines. While their spectral position is in the vicinity but not exactly coincident with the experimentally observed emission peak, no truly quantitative comparison is again possible due to the strong sensitivity on the pulse speed $v$ and the Bessel angle $\delta$. However, as the resonant radiation process is stimulated by the huge number of photons present in the pump pulse, we expect its emission to be fully coherent and its intensity to be far larger than the one of the spontaneous radiation.

As last competing effects, it must be mentioned that the nonstationarity of the pulse might be responsible for other nontrivial classical or quantum~\cite{angus} emission processes during the pulse propagation.
 
In future experiments, an unambiguous signature of the spontaneous emission nature of the emission may be obtained from the nonclassical positive correlations between pairs of modes with opposite comoving frequencies: As it was discussed in the context of analog models based on Bose-Einstein condensates, the spontaneous radiation is expected to violate Cauchy-Schwartz-like inequalities~\cite{zapata,chris} and the emitted photons are expected to consist of strongly entangled pairs~\cite{entanglement}.

\section{Conclusions}

The main conclusion of this article is the recognition of the fundamental role played by the material dispersion in determining the spectral properties of spontaneous vacuum emission from analog white holes in nonlinear optical systems.

In nonpolar dielectrics like diamond, where the low-energy dispersion of photons is linear and a description in terms of quantum field theory in curved spacetime is legitimate, we confirm the known result~\cite{gradino,faccionuovo} that the Hawking emission seen from the comoving frame has a thermal-like shape. When moving to the laboratory frame, the emission turns out to be peaked at the wavelength $\Lambda_0$ at which the pulse speed equals the phase velocity of light.

In other materials, like the fused silica used in the experiments~\cite{faccioexp,faccionjp}, the emission spectrum is significantly modified as the white-hole horizon is only active within a finite frequency window. As a result, the thermal character of the comoving frame emission is lost. Doppler transformation to the laboratory frame then leads to a sizable shift of the emission peak from the expected wavelength $\Lambda_0$, which is altogether more important as the refractive index modulation is weak.

As an illustrative example, we have performed a critical comparison with experimental data of~\cite{faccioexp,faccionjp}. While the theoretical spectrum of the spontaneous emission is  in good qualitative agreement with the measured one, the strong sensitivity of the spectral features on experimental parameters such as the pulse speed makes a quantitative comparison very difficult. In particular, we are not able to rule out competing effects with similar emission spectra and possibly stronger intensities such as negative resonant radiation. An incontrovertible proof of the spontaneous nature of the emission could be obtained by looking at the nonclassical correlations among the emitted phonons.

\begin{acknowledgments}

We are grateful to D. Faccio for continuous exchanges on the experiment and to R. Balbinot, T. Jacobson, U. Leonhardt, S. Liberati, R. Parentani, and all participants in the workshop on Effective Gravity in Fluids and Superfluids at the International Center for Theoretical Physics, Trieste, for stimulating discussions. We also thank D. Faccio and R. Parentani for their comments on the manuscript.  This work has been supported by European Research Council through the QGBE grant.

\end{acknowledgments}

\appendix

\section{Short review of the formalsm}

\subsection{Field equation and mode expansion}
\label{app:equations}

In this Appendix, we briefly review the formalism developed in~\cite{gradino} to describe light propagation in analog white-hole configurations and show how the rate of spontaneously emitted particles is computed.
In the laboratory reference frame, the one-dimensional Lagrangian density of the electromagnetic field coupled to $N$ polarization fields $P_i$ is
\begin{equation}\label{eq:lagrangian}
 {\cal L}_l=\frac{{(\partial_{T} A)}^2}{8\pi c^2}-\frac{{(\partial_{X} A)}^2}{8\pi}+\sum_{i=1}^N\left[\frac{{(\partial_{T} P_i)}^2}{2\beta_i\Om_i^2}-\frac{P_i^2}{2\beta_i}+\frac{1}{c}A\,{\partial_{T} P_i}\right].
\end{equation}
After a Lorentz boost, the Lagrangian density in the reference frame comoving with the pulse at velocity $v$ is
\begin{multline}\label{eq:lagrangianboost}
 {\cal L}=\frac{{\dot A}^2}{8\pi c^2}-\frac{{A'}^2}{8\pi}
 +\sum_{i=1}^3\left[
 \frac{\gamma^2}{2\beta_i\Om_i^2}\left({\dot P}_i-vP_i'\right)^2\right.\\\left.
-\frac{P_i^2}{2\beta_i}+\frac{\gamma}{c}A\left({\dot P_i}-vP_i'\right)^2\right],
\end{multline}
where dot and prime denote derivation with respect to the comoving time and space coordinates, $t$ and $x$, respectively.

As usual, the conjugate momenta are obtained by varying the Lagrangian
\begin{equation}
 L=\int \dx\,{\cal L}
\end{equation}
with respect to the time derivatives of $A$ and $P_i$:
\begin{equation}\label{eq:momenta}
 \Pi_A=\frac{\dot A}{4\pi c^2},\qquad\Pi_{P_i}=\frac{\gamma^2}{\beta_i\Om_i^2}\left({\dot P}_i-vP_i'\right)+\frac{\gamma}{c}A.
\end{equation}
The fields $A$, $P_i$, $\Pi_A$, and $\Pi_{P_i}$ satisfy the following equations:
\begin{align}
 &\dot A=4\pi c^2\,\Pi_A,\label{eq:system1}\\
 &\dot P_i=\frac{\beta_i\Om_i^2}{\gamma^2}\left(\Pi_{P_i}-\frac{\gamma}{c}A\right)+vP_i',\label{eq:system2}\\
 &{\dot \Pi}_A=\frac{A''}{4\pi}+\sum_{i=1}^3\left[\frac{\beta_i\Om_i^2}{\gamma c}\left(\Pi_{P_i}-\frac{\gamma}{c}A\right)\right],\label{eq:system3}\\
 &{\dot \Pi}_{P_i}=-\frac{P_i}{\beta_i}+\partial_{x}\left(v\Pi_{P_i}\right).\label{eq:system4}
\end{align}

Defining the eight-dimensional vector
\begin{equation}
 V=
 \begin{pmatrix}
  A & P_1 & P_2 & P_3 & \Pi_A & \Pi_{P_1} & \Pi_{P_2} & \Pi_{P_3}
 \end{pmatrix}^T
\end{equation}
and the matrix
\begin{equation}
 \eta=
 \begin{pmatrix}
  0&I_4\\-I_4&0
 \end{pmatrix},
\end{equation}
($I_4$ is the $4\times4$ identity matrix), the Hamilton equations can be written in a compact form as
\begin{equation}\label{eq:compact}
 \dot V=\eta(\nabla_V {\cal H}).
\end{equation}
Furthermore, a conserved scalar product can be defined on the space of the solutions of the above equation of motion:
\begin{equation}\label{eq:scalar}
 \langle V_1,V_2\rangle=\frac{\ii}{\hbar}\int \dx\, V_1^\dagger(x,t)\,\eta\, V_2(x,t).
\end{equation}

Being the system stationary in the reference system comoving with the pulse, it is convenient to expand the real field $V$ on a basis of frequency eigenmodes $V_\omega$, rather than, as usually done, on a basis of wavevector eigenmodes:
\begin{equation}\label{eq:expansion}
 V=\int
 \dd
\om\sum_\alpha \left(V_{\om}^{\alpha}{\hat a}_{\om}^\alpha+V_{\om}^{\alpha*}{\hat a}_{\om}^{\alpha\dagger}\right),
\end{equation}
where
\begin{equation}\label{eq:a}
 \hat a_\om^\alpha=\langle V_\om^\alpha,V\rangle.
\end{equation}
Here, the label $\alpha$ denotes various modes with the same eigenfrequency $\om$, $V_{\om}^{\alpha}$ are properly normalized with respect to the norm induced by the scalar product defined in Eq.~\eqref{eq:scalar}, and the integral in Eq.~\eqref{eq:expansion} generally includes both positive and negative frequency modes with positive norm.

In the asymptotic regions far from the pulse edge, the system is homogeneous and the parameters $\Om_i$, $\beta_i$, and $v$ are constant both in time and space. In this situation one can chose $V_{\om}^\alpha$ as momentum eigenmodes:
\begin{equation}\label{eq:momentumeigenmode}
 V_{\om}^{\alpha}(x,t)=\ee^{-\ii\om t+\ii k_\alpha x} \bar V_{\om}^{\alpha},
\end{equation}
where $\bar V_{\om}^{\alpha}$ is a vector of constant $\mathbb{C}$ numbers, satisfying
\begin{equation}\label{eq:fouriersystem}
 -\ii\om\bar V_{\om}^{\alpha}=\eta\, {\cal K}(k_\alpha)\,\bar V_{\om}^{\alpha},
\end{equation}
where
\begin{widetext}
\begin{equation}
 {\cal K}(k_\alpha)=
 \begin{pmatrix}
  k_\alpha^2/4\pi+\sum_{i=1}^3\beta_i\Om_i^2/c^2  & 0 & 0 & 0 & 0 & -\beta_1\Om_1^2/\gamma c & -\beta_2 \Om_2^2/\gamma c & -\beta_3\Om_3^2/\gamma c\\
  0 & 1/\beta_1 & 0 & 0 & 0 & -\ii k_\alpha v & 0 & 0 \\
  0 & 0 & 1/\beta_2 & 0 & 0 & 0 & -\ii k_\alpha v & 0 \\
  0 & 0 & 0 & 1/\beta_3 & 0 & 0 & 0 & -\ii k_\alpha v \\
  0 & 0 & 0 & 0 & 4\pi c^2 & 0 & 0 & 0 \\
  -\beta_1\Om_1^2/\gamma c & + \ii k_\alpha v & 0 & 0 & 0 & \beta_1\Om_1^2/\gamma^2 & 0 & 0 \\
  -\beta_2\Om_2^2/\gamma c & 0 & + \ii k_\alpha v & 0 & 0 & 0 & \beta_2\Om_2^2/\gamma^2 & 0 \\
  -\beta_3\Om_3^2/\gamma c & 0 & 0 & + \ii k_\alpha v & 0 & 0 & 0 & \beta_3\Om_3^2/\gamma^2\\
 \end{pmatrix}.
\end{equation}
\end{widetext}
The eigenvalues of the above matrix determine the dispersion relation
\begin{equation}\label{eq:dispersion}
 c^2 k_\alpha^2 = \om^2 +\sum_{i=1}^3\frac{4\pi\beta_i\, \gamma^2(\om+v k)^2}{1-\gamma^2(\om+v k)^2/\Om_i^2},
\end{equation}
which has the form of the Sellmeier dispersion relation in the comoving frame.
The eigenvectors corresponding to the solutions of this dispersion relation can be normalized using the above-introduced scalar product. Since the scalar product is not positive definite, there exist modes with negative norm. On one hand, it is possible to show that all negative norm modes have a negative laboratory frequency $\Om=\gamma(\om+v k)$; on the other hand, for a given positive comoving frequency $\om$ there exist both positive and negative norm modes. This fact is responsible for the emission of quantum vacuum radiation.

This implies that, in the expansion of the field $V$, the Fock operators associated with those positive-$\om$ modes are not destruction but instead creation operators. Naming $P$ the set of positive norm modes $V_{\om}^{\alpha}$, labeled by $\alpha$, and $N$ the set of negative norm modes $V_{\om}^{\tilde\alpha}$, labeled by $\tilde\alpha$, $V$ becomes
\begin{multline}\label{eq:expansionadagger}
 V=\int_0^\infty\dd
\om\,\ee^{-\ii\om t}\left(\sum_{\alpha\in P} \ee^{+\ii k_\alpha x} \bar V_{\om}^{\alpha}{\hat a}_{\om}^\alpha
\right.\\\left.
+\sum_{\tilde\alpha\in N}\ee^{+\ii k_{\tilde\alpha} x} \bar V_{\om}^{\tilde\alpha}
{\hat a}_{\om}^{\tilde\alpha\dagger}\right)+\mbox{H.c.},
\end{multline}
where H.c. stands for Hermitian conjugate. Note that the positive frequency part of the field (i.e., evolving with $\ee^{-\ii\om t}$) mixes creation $\hat a_\om^\alpha$ and destruction $\hat a_{\om}^{\tilde\alpha\dagger}$ operators.

In an analog white-hole geometry, obtained by pasting together two homogeneous regions as described in Sec.~\ref{sec:theory}, a frequency eigenmode $V_\om^\alpha$ [see Eq.~\eqref{eq:expansion}] can be written as
\begin{equation}\label{eq:leftright}
 V_\om^\alpha=\sum_{\alpha}L^{\alpha}_\om\, V_{\om,L}^{\alpha}\,\theta(-x)
 +\sum_{\alpha}R^{\alpha}_\om\, V_{\om,R}^{\alpha}\,\theta(x),
\end{equation}
where $L^{\alpha}_\om$ and $R^{\alpha}_\om$ are constant, and $V_{\om,L}^{\alpha}$ and $V_{\om,R}^{\alpha}$ are
frequency-momentum eigenmodes, as in Eq.~\eqref{eq:momentumeigenmode}; that is, they are
solutions of the field equation~\eqref{eq:compact} in the homogeneous left ($x<0$) and right ($x>0$) regions, respectively.
The relations between $L^{\alpha}_\om$ and $R^{\alpha}_\om$ are determined by solving the field equation in a neighborhood of the pulse edge at $x=0$.

\subsection{Scattering modes}
\label{app:modes}

\begin{figure*}
\begin{center}
\includegraphics[width=0.86\textwidth]{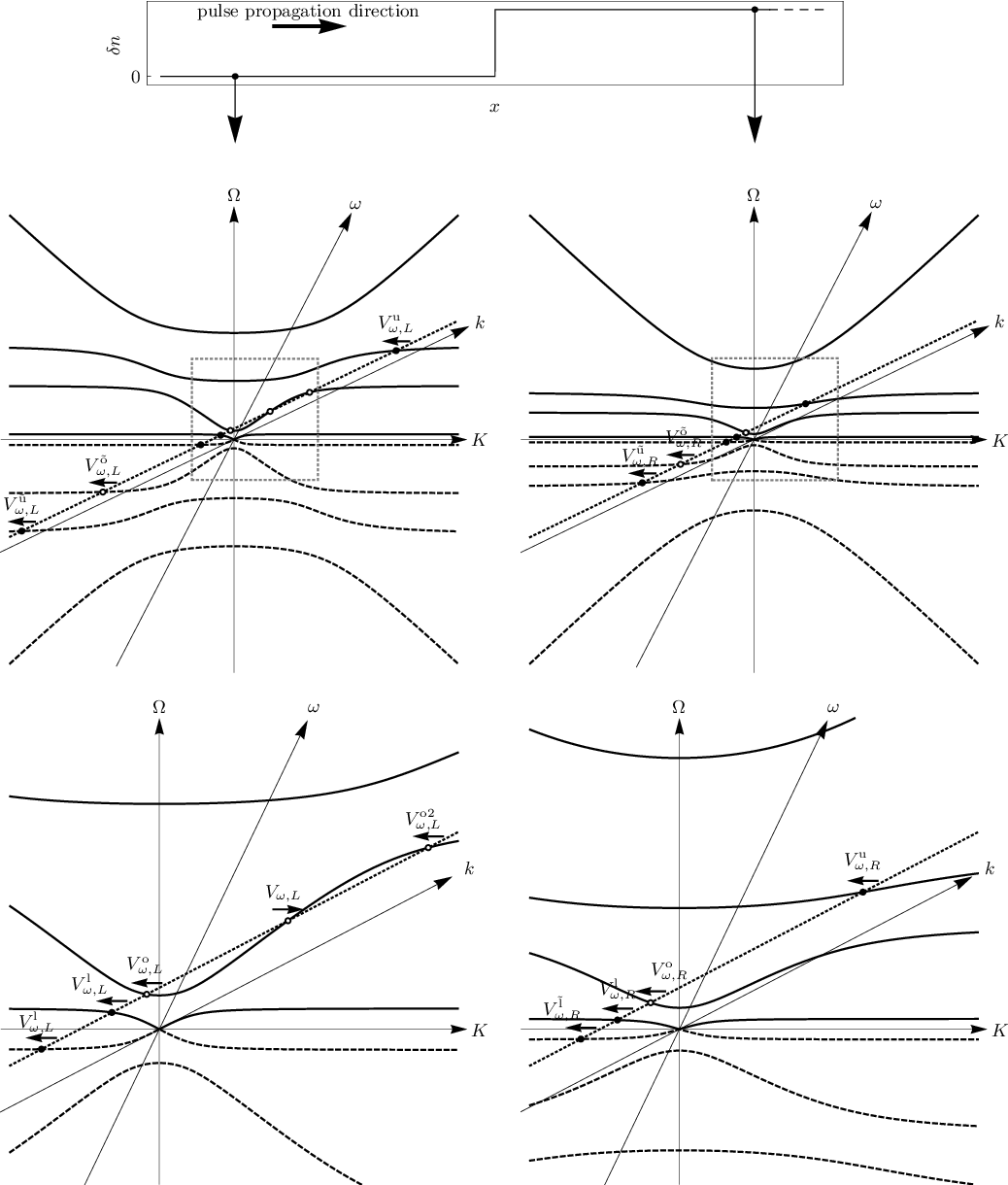}
\end{center}
\caption{Graphical representation of the Sellmeier dispersion relation, as seen from the laboratory reference frame $(\Om,K)$, for $x<0$ (left panels) and $x>0$ (right panels).
The $(\om,k)$ axes of the comoving reference frame are obtained through a boost of velocity $v$.
As sketched in the top panel, the refractive index in the right region is larger than in the left region. This difference in the refractive index is obtained by properly changing the parameters $\beta_i$ and $\Om_i$ appearing in the Lagrangian [Eq.~\eqref{eq:lagrangian}].
In this plot the values of the velocity $v$ and of the refractive index change $\delta n$ have been arbitrarily chosen for illustrative purposes.
The bottom panels are enlargements of the gray dot-bordered squared of the respective upper panels. The dispersion relation is graphically solved for a fixed comoving frequency $\om$, chosen in the frequency window in which the black-hole horizon is present.
Solutions appear both on the positive norm positive-$\Om$ branches (solid curves) and on the negative norm negative-$\Om$ branches (dashed curves).
The empty dots denote solutions on the optical branches with positive (o) and negative ($\rm\tilde o$) frequency $\Om$. The arrows indicate the direction of propagation (group velocity in the comoving frame) of the associated modes $V_{\om,L/R}^{\alpha/\tilde\alpha}$. In the right region (right panels), the dispersion relation has only six real-$k$ solutions. 
In the left region (left panels), the number of real-$k$ solutions is eight, and the two extra solutions are one left-going ($V_{\om,L}^{\rm o2}$) and one right-going ($V_{\om,L}$).}
\label{fig:dispersion}
\end{figure*}

In Fig.~\ref{fig:dispersion} the full Sellmeier dispersion relation is plotted for $x<0$ (upper left panel) and $x>0$ (upper right panel) in the laboratory reference frame $(\Om,K)$ [see Eq.~\eqref{eq:sellmeier}]. A boost is then performed on the axes and the new axes $(\om,k)$ in the reference frame comoving with the pulse are drawn [see Eq.~\eqref{eq:dispersion}]. The central region of those plots (gray dot-bordered square) is enlarged in the bottom panels.
The dispersion relation is graphically solved for a fixed value of the comoving frequency $\om$ (dashed line), with $\ommin<\om<\ommax$. There are eight branches: four with positive laboratory frequency $\Om$ (solid curves) and four with negative $\Om$ (dashed curves), symmetrically placed in the lower plane. In this configuration $\om$ is small enough that no solution belongs to the highest (positive or negative) energy branches.
We therefore name only the six branches of the dispersion relation with low energy $|\Om|$. In the upper plane ($\Om>0$), starting from the lowest-energy branch, we call them lower (l), optical (o), and upper (u). Symmetrically, the three branches with negative laboratory frequency $\Om$ and negative norm are labeled by $\rm \tilde l$, $\rm \tilde o$, and $\rm \tilde u$.
Accordingly, the solutions of the dispersion relation are labeled by a superscript l, o, u, $\rm \tilde l$, $\rm \tilde o$, and $\rm \tilde u$.
The solutions on the positive and negative frequency optical branches and $\rm\tilde o$ are denoted by an empty dot. The arrow above each solution indicates the direction of propagation (group velocity in the comoving frame) of the associated mode $V_{\om,L/R}^{\alpha/\tilde\alpha}$.

Note that, for $x>0$ (right panels), there are six real-$k$ solutions, all corresponding to left-going modes. The remaining two solutions of Eq.~\eqref{eq:dispersion} have complex conjugate $k$. They are associated with exponentially growing ($V_{\om,R}^{\rm grow}$) and decaying ($V_{\om,R}^{\rm dec}$) modes for $x\to\infty$.
For $x<0$ (left panels), instead, the eight solutions are all real.
The two extra real solutions, that do not have a corresponding solution on the right side, belong to the optical branch and are associated, respectively, with a left-going mode, named $V_{\om,L}^{\rm o2}$, and with the unique right-going mode, simply named $V_{\om,L}$, without any superscript.

By combining those asymptotic plane-wave modes propagating in the flat left and right regions, two relevant bases of globally defined asymptotically bounded modes (GDMs) (not diverging at infinity) can be constructed.

We define the \emph{in} basis as the set of \emph{in} modes, whose asymptotic decomposition [Eq.~\eqref{eq:leftright}] has only one AM with group velocity $v_g$ directed toward $x=0$. We say that the group velocity of an AM is directed toward the horizon if $v_g>0$ ($v_g<0$) for modes which are solutions of the mode equation in the left (right) region.

Analogously, we define the \emph{out} basis as the set of \emph{out} modes, whose asymptotic decomposition has only one AM with group velocity directed toward $x=-\infty$ ($x=+\infty$), if the AM is a solution of the field equation in the left (right) region.

As in Eq.~\eqref{eq:expansionadagger}, the field operator $V$ is expanded indifferently with respect either to the \emph{in} or to the \emph{out} basis of GDMs.
\begin{align}\label{eq:expansiondagger}
 V
 &=\int_0^\infty\!\!\!\dd
\om\,\ee^{-\ii\om t}\left(\sum_{\alpha\in P}  V_{\om}^{{\rm in},\alpha}{\hat a}_{\om}^{{\rm in},\alpha}
+\sum_{\tilde\alpha\in N} V_{\om}^{{\rm in},\tilde\alpha}
{\hat a}_{\om}^{{\rm in},\tilde\alpha\dagger}\right)\nonumber\\
&\qquad\qquad
+\mbox{H.c.}\\
&=\int_0^\infty\!\!\!\dd
\om\,\ee^{-\ii\om t}\left(\sum_{\alpha\in P} V_{\om}^{{\rm out},\alpha}{\hat a}_{\om}^{{\rm out},\alpha}
+\sum_{\tilde\alpha\in N} V_{\om}^{{\rm out},\tilde\alpha}
{\hat a}_{\om}^{{\rm out},\tilde\alpha\dagger}\right)
\nonumber\\&\qquad\qquad
+\mbox{H.c.}.
\end{align}

The transformation between the two bases follows straightforwardly from the construction of those bases. For instance the incoming negative frequency mode $V_{\omega}^{\rm in,\tilde o}$ (the global defined mode whose unique asymptotic branch with group velocity directed toward the horizon is $V_{\omega}^{R,\tilde o}$) is
\begin{multline}
 V_{\omega}^{\rm in,\tilde o}=
A_{\omega}^{\rm o,l}V_{\omega}^{\rm out,\tilde l}+
\alpha_\om V_{\omega}^{\rm out,\tilde o}+
A_\om^{\rm o,u}V_{\omega}^{\rm out,\tilde u}\\
+B_{\omega}^{\rm o,l} V_{\omega}^{\rm out,l}+
B_{\omega} V_{\omega}^{\rm out,o}+
B_{\omega}^{\rm o,u} V_{\omega}^{\rm out,u}+
\beta_\om V_{\omega}^{\rm out}.
\end{multline}
Repeating this procedure for each \emph{in} mode, the scattering matrix $S$ is fully determined:
\begin{equation}
 V_{\om}^{\rm in,\beta}=\sum_{\beta'} S^{\beta\beta'}V_{\om}^{\rm out,\beta'}.
\end{equation}

Once the scattering matrix is known, the rate per unit time and unit bandwidth of spontaneous emission on a certain positive norm mode is simply given by the sum of the squared absolute values of the amplitudes of the matrix elements relating all the incoming negative norm modes and to the chosen outgoing positive norm mode.

\bibliography{kinetics}

\end{document}